\documentclass[12pt]{article}
\usepackage{epsfig}
\begin{document}
\begin{center}
{\Large \bf 
Evaluation of QCD sum rules for HADES}\footnote{
Talk at the XL International Winter Meeting on
Nuclear Physics, Bormio (Italy), January 20 - 26, 2002.
Supported by BMBF 06DR921, GSI and DAAD.}
\vskip 5mm
{\sc S.~Zschocke$^{\, a)}$, \underline{B.~K\"ampfer}$^{\, a)}$, 
O.P.~Pavlenko$^{\, a, b)}$,
Gy. Wolf$^{\, c)}$}

\vskip 5mm

$^{a)}$ {Forschungszentrum Rossendorf, PF 510119, 01314 Dresden, Germany}\\
$^{b)}$ {Institute for Theoretical Physics, 252143 Kiev - 143, Ukraine}\\
$^{c)}$ {RKMI - KFKI Budapest, Hungary}
\end{center}
\vskip 5mm

\begin{center}
\begin{minipage}{130mm}
{\small
QCD sum rules are evaluated at finite nucleon densities and temperatures 
to determine the change of pole mass parameters
for the lightest vector mesons $\rho$, $\omega$ and $\phi$ in a  
strongly interacting medium at conditions relevant for the starting
experiments at HADES.
The role of the four-quark condensate is highlighted.
A few estimates (within a fire ball model and BUU calculations)
of dilepton spectra in heavy-ion collisions at 1 AGeV
are presented.
}
\end{minipage}
\end{center}

\section{Introduction}

Hadrons are excitations of the QCD vacuum.
There are many different species and masses of them.
Their excitation spectra resemble, to some extent, 
to level schemes of atoms.
The latter ones are known to be modified by external fields: 
The Stark and Zeeman effects manifest themselves as shifts of spectral lines.
By analogy, one might expect a similar behavior of hadrons 
in strongly interacting matter. In a first approximation, 
hadrons should experience mass shifts, but considerable
broadenings or strong modifications (e.g. mixing) 
of the spectral strengths are conceivable, too.
A verification and understanding of such in-medium modifications
of hadrons would allow to contribute to the 
''mystery of particle masses''. 

This topic was essentially
triggered by the Brown-Rho scaling hypothesis \cite{BR}
according to which the masses of the light vector mesons
($V = \rho, \omega, \phi$) behave as
$m_V^2 \propto \langle \bar q q \rangle$, where 
$\langle \bar q q \rangle$ is the chiral condensate. In such a way
fundamental and non-perturbative QCD quantities can be accessible.
This idea influenced strongly the 
HADES project which aims at measuring the density and temperature
dependence of 
$\langle \bar q q \rangle$ via vector meson 
''spectral lines''. A strong impact on the field had
QCD sum rules \`a la Shifman, Vainshtein, Zakharov \cite{SVZ}
which, indeed, correlate condensates (and moments of the parton
distributions in hadrons) and the pole masses of hadrons.

The purpose of the present contribution is two-fold.
First, we revisit the QCD sum rule approach to estimate the change
of the vector meson pole mass parameter
at finite baryon density $n$ \underline{and} temperature $T$.
Second, we present dynamical calculations, basing on 
hydrodynamics (fire ball model) and transport calculations
(BUU) to elucidate the prospects to verify in the starting
experiments at HADES the predicted in-medium modifications
of vector mesons. 

\section{QCD sum rules revisited}

The basic object is the retarded current-current correlation function
\begin{equation}
\Pi^R_{\mu \nu} (q; \mu_N, T) =
i \int d^4 x \;   
{\rm e}^{i \, q  x} \, \Theta (x^0) \,
\langle [J_{\mu}(x), J_{\nu}(0)] \rangle_{\mu_N, T},\\
\end{equation}
with conserved vector currents  
$J_\mu^{\rho (\omega), \phi} 
= \frac 12 ( \bar u \gamma_\mu u \mp \bar d \gamma_\mu d)$,
$\bar s \gamma_\mu s$
carrying the quantum numbers of the $\rho$, $\omega$ and $\phi$
meson, respectively, but expressed by quark fields. 
$\langle \cdots \rangle_{\mu_N, T}$
means grand canonical averaging with respect to chemical
potential $\mu_N$ and temperature $T$.
Here we focus on vector mesons at rest and constrain ourselves
on the longitudinal part $\Pi^R_L$ of the correlator $\Pi^R_{\mu \nu}$.

The sum rule approach rests on a comparison of two different evaluations
of $\Pi^R_L$. One can use (i) the analyticity properties
to note a dispersion relation  
\begin{equation}
\Pi_{L}^{R} (q^0; \mu_N , T)
= \frac 1 \pi  
\int_0^\infty d s \; 
\frac{{\rm Im} \Pi^R_L (s; \mu_N , T)}{s - (q_0 + i \epsilon)^2}
+ \cdots,
\end{equation}
where $\cdots$ indicate subtractions, and (ii) the 
Operator Product Expansion at large $Q^2 \equiv - q_0^2 > 0$
\begin{equation}
\Pi^R_L (Q^2) = 
- C_0 \; \ln Q^2 
+ \sum_{n=1}^{\infty} \frac{C_n}{Q^{2n}},   
\end{equation}
where  $C_N$ are quantities containing Wilson coefficients 
and condensates and moments of parton distributions.
Defining the spectral density 
$\rho_{\rm had}(s; \mu_N , T) 
= \frac 1 \pi {\rm Im} \Pi^R_L (s; \mu_N , T)$ 
and performing a Borel transformation one arrives at the
QCD sum rule (QSR)
\begin{equation}
\int_0^\infty d s \; \rho_{\rm had}(s) \, \mbox{e}^{-s / M^2} =
M^2 \left( C_0 + \sum_{n=1}^{\infty} \frac{C_n}{(n -1)! \, M^{2n}} 
\right),
\label{QSR}
\end{equation}
where $M$ is the Borel mass.
Assuming that the r.h.s. is calculable from first principles
one can use the QSR
(i) either as consistency check of hadronic models for
$\rho_{\rm had}(s)$ (cf.\ \cite{Klingl_Weise}), or
(ii) as one equation for a physically relevant parameter
in a simple parameterization of $\rho_{\rm had}(s)$, or
(iii) as constraint for a few parameters entering models
of $\rho_{\rm had}(s)$ (e.g. to find a width-mass relation
\cite{Leupolt_Mosel}, or to interrelate various parameters
with the chiral gap \cite{Weise}).

We follow here the possibility (ii) as exercised by many
previous authors like in \cite{SVZ,Hatsuda}.
Inspired by the experimental results on the cross sections
of the reactions $e^+ e^- \to n \pi, K^+ K^-$, which is
intimately related to $\rho_{\rm had}$ and which exhibits
one low-lying distinctive resonance (either $\rho$ or
$\omega$ or $\phi$) followed by a flat continuum, one
parameterizes 
\begin{equation}
\rho^V_{\rm had} (s; T, n) = 
F_V \; \delta (s - m_V^2) + C_0 \Theta (s - s_0^V)
+ (\rho_{\rm scatt}^{V, \pi} + \rho_{\rm scatt}^{V, N}) \delta(s),
\end{equation}
where $m_V$ is the wanted pole mass parameter, 
$s^V_0$ stands for the continuum threshold,
and $\rho_{\rm scatt}^{V, \pi (N)}$ denote the forward scattering 
amplitudes or Landau damping terms, which read for the $\rho$
meson
$\rho_{\rm scatt}^{\rho, N} =
(48 \pi^2)^{-1} \int_{4 M_N^2}^{\infty} 
d \omega^2 \, \hat n_F
\sqrt{1 - 4 M_N^2 / \omega^2} 
\left[ 2 + 4 M_N^2 / \omega^2 \right]$,
$\rho_{\rm scatt}^{\rho, \pi} =
(24 \pi^2)^{-1} \int_{4 m_{\pi}^2}^{\infty}
d \omega^2 \, \hat n_B$
$\sqrt{1 - 4 m_{\pi}^2 / \omega^2} 
\left[ 2 + 4 m_{\pi}^2/ \omega^2 \right]$
with
$\hat n_F= [{\rm e}^{(\omega - 2 \mu_N)/2 T} + 1]^{-1}$,
$\hat n_B = [{\rm e}^{\omega/2 T} - 1]^{-1}$,
while for the $\omega$ meson, $\rho_{\rm scatt}^{\omega, \pi} = 0$ 
due to symmetry,
and $\rho_{\rm scatt}^{\omega, N} = 9 \rho_{\rm scatt}^{\rho, N}$
due to the different isospin structure of the $\rho$  and $\omega$ mesons
\cite{Abee};
for the $\phi$ meson the Landau damping for the pion and nucleon gas
is negligible \cite{Asakawa_Ko}.

Truncating the r.h.s of eq.~(\ref{QSR})
the sum rule might be cast in the form
\begin{equation}
m_V^2 = 
M^2 \frac{C_0 \left( 1 - [1 + \frac{s_0^V}{M^2}] {\rm e}^{-s_0^V/M^2} 
\right) 
- C_2 M^{-4} - C_3 M^{-6}}
{ C_0 \left( 1 - {\rm e}^{-s_0^V/M^2} \right)
+ C_1 M^{-2}     
+ C_2 M^{-4} + \frac12 C_3 M^{-6} 
- (\rho_{\rm scatt}^{V, N} + \rho_{\rm scatt}^{V, \pi} ) M^{-2}},
\label{m_V}
\end{equation}
where the coefficients up to mass dimension 6 have been calculated
by many authors in the low-temperature or low-density approximation.
In the low-$T$ \underline{and} low-$n_N$ approximation our result is
\begin{eqnarray}
C_0
& = & 
\frac{1}{8 \pi^2} \left( 1 + \frac{\alpha_s(\mu^2)}{\pi} \right),
\quad 
C_1 = 
- \frac{3 m_q^2}{4 \pi^2},
\label{eq_5.1}\\ 
C_2 & = & q_2 + g_2 + a_2, 
\quad \quad \quad \quad
C_3 =
q_4 + a_4, \label{eq_28}\\
q_2 & = & m_q \langle \bar u u \rangle_0 
+ 2 M_N \sigma_N Y I_1^N + \frac34 m_\pi^2 \xi^{\rho (\omega)} I_1^\pi 
\label{eq_q2}\\
g_2 & = & \frac{1}{24} \langle \frac{\alpha_s}{\pi} G^2 \rangle_0 
- \frac{4}{27} M_N M_N^0 I_1^N 
- \frac{1}{18} m_\pi^2 I_1^\pi \\  
q_4 & = & - \frac{112}{81} \pi \alpha_s \kappa_0 
\langle \bar u u \rangle_0^2 
\left[ 1 -
\frac{\kappa_N}{\kappa_0} 
\frac{4 M_N \sigma_N}{m_q (-\langle \bar u u \rangle_0)} 
Y I_1^N  
- \frac{36}{7 f_\pi^2} \xi^\rho I_1^\pi \right] \label{eq_31}\\
a_2 & = & M_N^2 \, A_{2,N}^{u+d} \, I_1^N + \frac43 A_{2,N}^{u+d} \, I_2^N 
 + \frac34 m_\pi^2 \, A_{2,\pi}^{u+d}\, I_1^\pi 
 + A_{2,\pi}^{u+d} \, I_2^\pi \label{eq_32}\\
a_4 & = & - \frac53 M_N^4 \,A_{4,N}^{u+d} \, I_1^N 
- \frac{20}{3} M_N^2 \, A_{4,N}^{u+d} \, I_2^N 
- \frac{16}{3} A_{4,N}^{u+d}\, I_3^N \nonumber\\
& & - \frac54 m_\pi^4 \, A_{4,\pi}^{u+d} \, I_1^\pi 
- 5 m_\pi^2 \, A_{4,\pi}^{u+d} \, I_2^\pi - 4 A_{4,\pi}^{u+d} \, I_3^\pi,
\label{eq_a4}
\end{eqnarray}
where $\xi^{\rho (\omega)} = 1$, $Y = 1$ for $\rho$ ($\omega$) 
mesons, and elsewhere $\xi^{\cdots} = 0$.   
The sequence of replacements $m_q \to m_s$,
$\langle \bar u u \rangle_0 \to \langle \bar s s \rangle_0$
$ Y \to y m_s / m_q$, and
$A_{n,(N,\pi)}^{u+d} \to A_{n,(N,\pi)}^s$
holds for the $\phi$ meson. 
The above integrals are
$ 
I_n^N =  \int d^3 k [(2 \pi)^3 E_k]^{-1} k^{2n-2} \, n_F, 
$
$
I_n^{\pi}  =  \int d^3 p [(2 \pi)^3 E_p]^{-1} p^{2n-2} \, n_B,
$ 
where $n_B = [{\rm e}^{E_p/T} - 1]^{-1}$ 
and $n_F = [{\rm e}^{(E_k - \mu_N)/T} + 1]^{-1}$
are thermal Boson and Fermion distributions,
and $\mu_N$ is the chemical potential related to the
nucleon density $n_N$. 

Specific for our approach is the treatment of the
four-quark condensate for which 
we extend the widely used ground state saturation
assumption for the in-medium
four-quark condensate at $T = 0$ in the following way
\begin{equation}
\langle (\bar q \gamma_\mu \gamma_5 \lambda^a q)^2 \rangle_{\mu_N} = -
\langle (\bar q \gamma_\mu \lambda^a q)^2 \rangle_{\mu_N} =
\frac{16}{9} \kappa (n_N) \langle \bar q q \rangle_{\mu_N}^2,
\label{eq_18}
\end{equation}
where the density dependent factor $\kappa (n_N)$ is introduced to control
a deviation from the exact ground state saturation.
As pointed out in \cite{Klingl_Weise_Kaiser} 
$\kappa (n_N)$ reflects the contribution
from the scalar low-energy excitations of the ground state and seems to be
weekly dependent on $n_N$, so that one can use $\kappa (n_N) = const$
as first approximation. In this case the density dependence of the 
four-quark condensate appears only via the density dependence of the
quark condensate squared. In linear density approximation it is given by
$
\langle \bar q q \rangle_{\mu_N}^2 = \langle \bar q q \rangle_0^2 +
\langle \bar q q \rangle_0
\langle N \vert \bar q q \vert N \rangle \frac{n_N}{M_N}.
$
We go beyond this approximation and include here a possible
linear density dependence of $\kappa (n_N)$,
keeping in mind that still $\kappa (n_N) = \kappa_0$ at $n_N = 0$ in accordance
with the vacuum case.
In leading order our extended ansatz results in an additional density dependence
parameterized by a constant factor
$\kappa_N$. Our parameterization of the four-quark condensate
at $T = 0$ has then the form
\begin{equation}
\langle (\bar q \gamma_\mu \gamma_5 \lambda^a q )^2 \rangle_{\mu_N} = 
\frac{16}{9} \langle \bar q q \rangle_0^2 \, \kappa_0 \,
\left( 1 + \frac{\kappa_N}{\kappa_0}
\frac{ \langle N \vert \bar q q \vert N \rangle}{\langle \bar q q \rangle_0 }
\frac{n_N}{M_N} \right) .
\label{eq_20}
\end{equation}
The limit $\kappa_N = \kappa_0 = 2.36$ is
used in \cite{Klingl_Weise_Kaiser}, 
while the parameterization $\kappa_N = 1.4$ and $\kappa_0 = 3.3$ with
$\langle \bar q q \rangle_0 = (- 230 \mbox{MeV})^3$ corresponds
to the choice in \cite{Hatsuda_Lee_Shiomi}. 
Below we vary the poorly known parameter $\kappa_N$ to estimate the contributions
of the four-quark condensates to the QSR.
The needed ansatz for the nucleon matrix element of the scalar four-quark
condensate can be extracted from the direct comparison of our
parameterization in eq.~(\ref{eq_20}) and the general expression
for the condensates via the matrix elements (the latter ones to be taken
at $T = 0$) as
$ 
\langle N \vert (\bar q \gamma_\mu \gamma_5 \lambda^a q)^2 \vert N \rangle = 
\frac{32}{9} \langle \bar q q \rangle_0 \, 
\langle N \vert \bar q q \vert N \rangle \, \kappa_N.
$ 
Since this matrix element does not depend on particle momenta
and temperature it can be employed also 
for evaluating the four-quark condensates in the general
case with $T \ne 0$ and $\mu_N \ne 0$.

The remaining parameters are specified by
$\alpha_s =0.38$,
$m_q = 0.0055$ GeV,
$m_s = 0.130$ GeV,
$f_{\pi} = 0.093$ GeV,
$M_N = 0.938$ GeV,
$M_N^0 = 0.770$ GeV,
$\sigma_N = 0.045$ GeV,
$y = 0.22$,
$m_{\pi} = 0.138$ GeV,
$\langle \bar u u \rangle_0 = \langle \bar d d \rangle_0 = 
(-0.245 \,{\rm GeV})^3$,
$\langle \bar s s \rangle_0 = 0.8 \langle \bar u u \rangle_0$, 
$\langle \frac{\alpha_s}{\pi} G^2 \rangle_0 =
(0.33 \, {\rm GeV})^4$, 
$A_{2,N}^{u+d} = 1.02$,
$A_{4,N}^{u+d} = 0.12$,
$A_{2,\pi}^{u+d} = 0.97$,
$A_{4,\pi}^{u+d} = 0.255$,
$A_{2,N}^{s} = 0.1$,
$A_{4,N}^{s} = 0.004$,
$A_{2,\pi}^{s} = 0.08$,
$A_{4,\pi}^{s} = 0.008$.

\begin{figure} 
\begin{minipage}[t]{9cm}
~\vskip 6mm
\epsfig{file=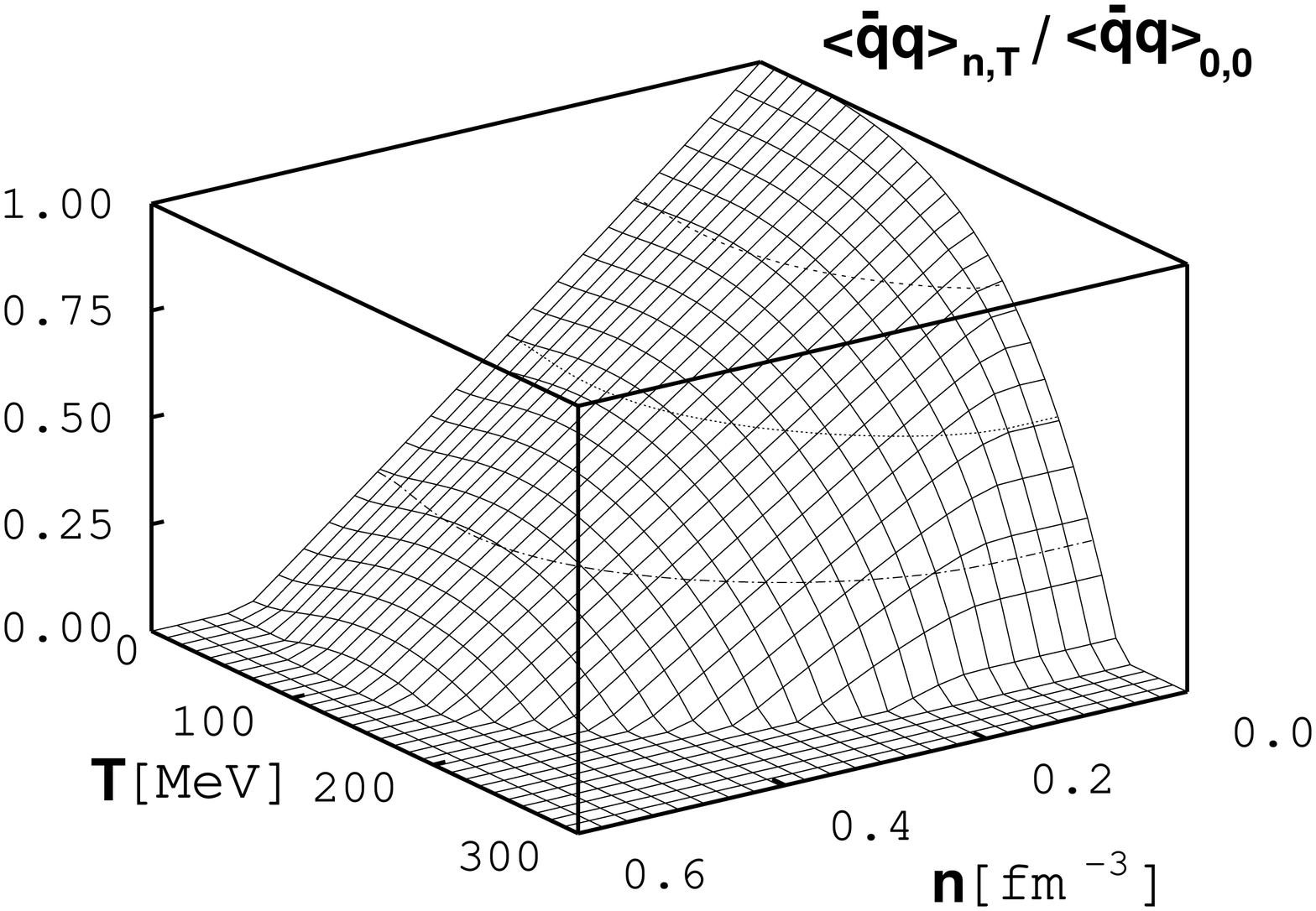,width=6cm,angle=-90}
\end{minipage}
\hspace*{-3mm}
\begin{minipage}[t]{5cm}
\epsfig{file=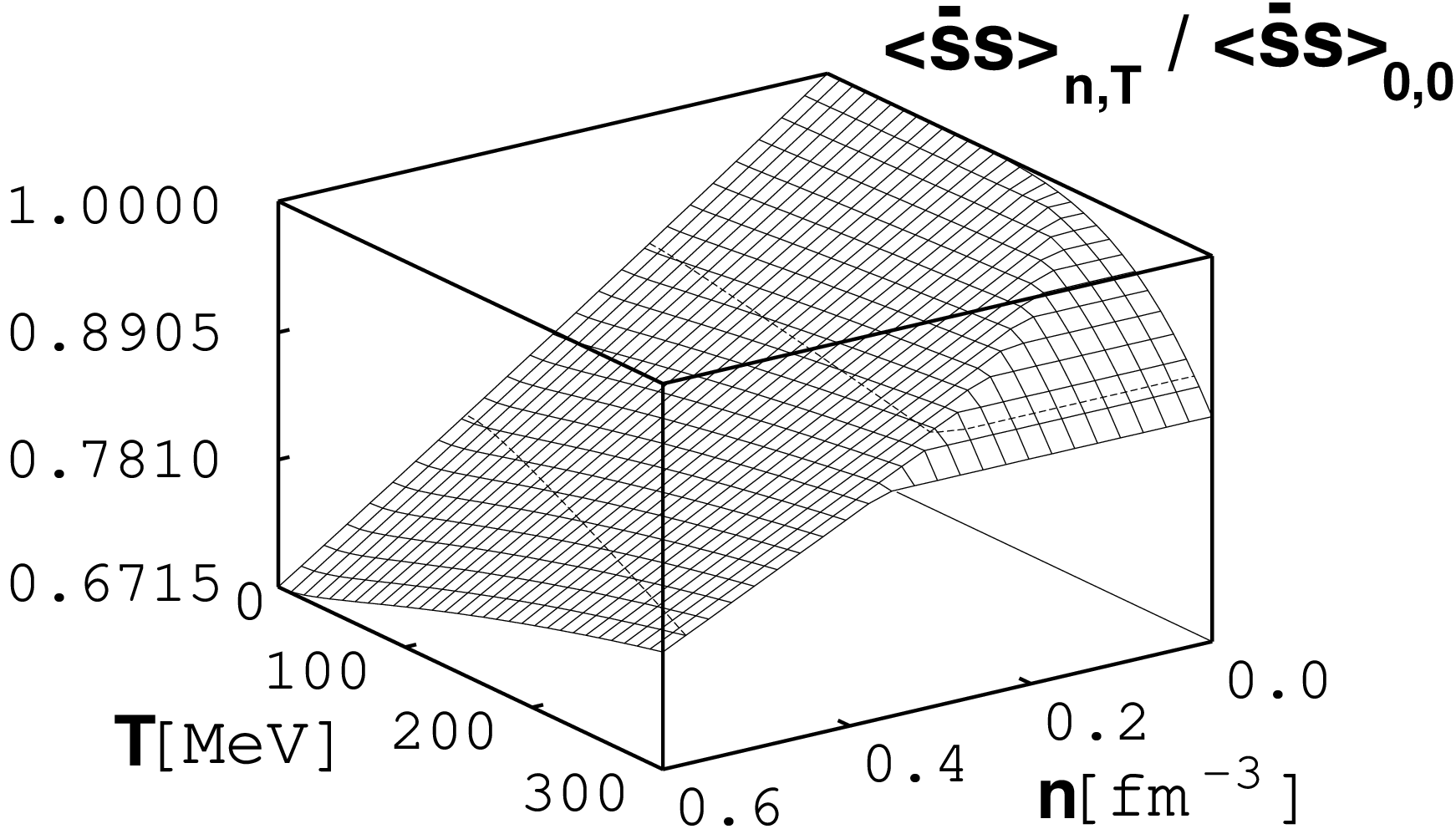,width=4cm,angle=-90}
\epsfig{file=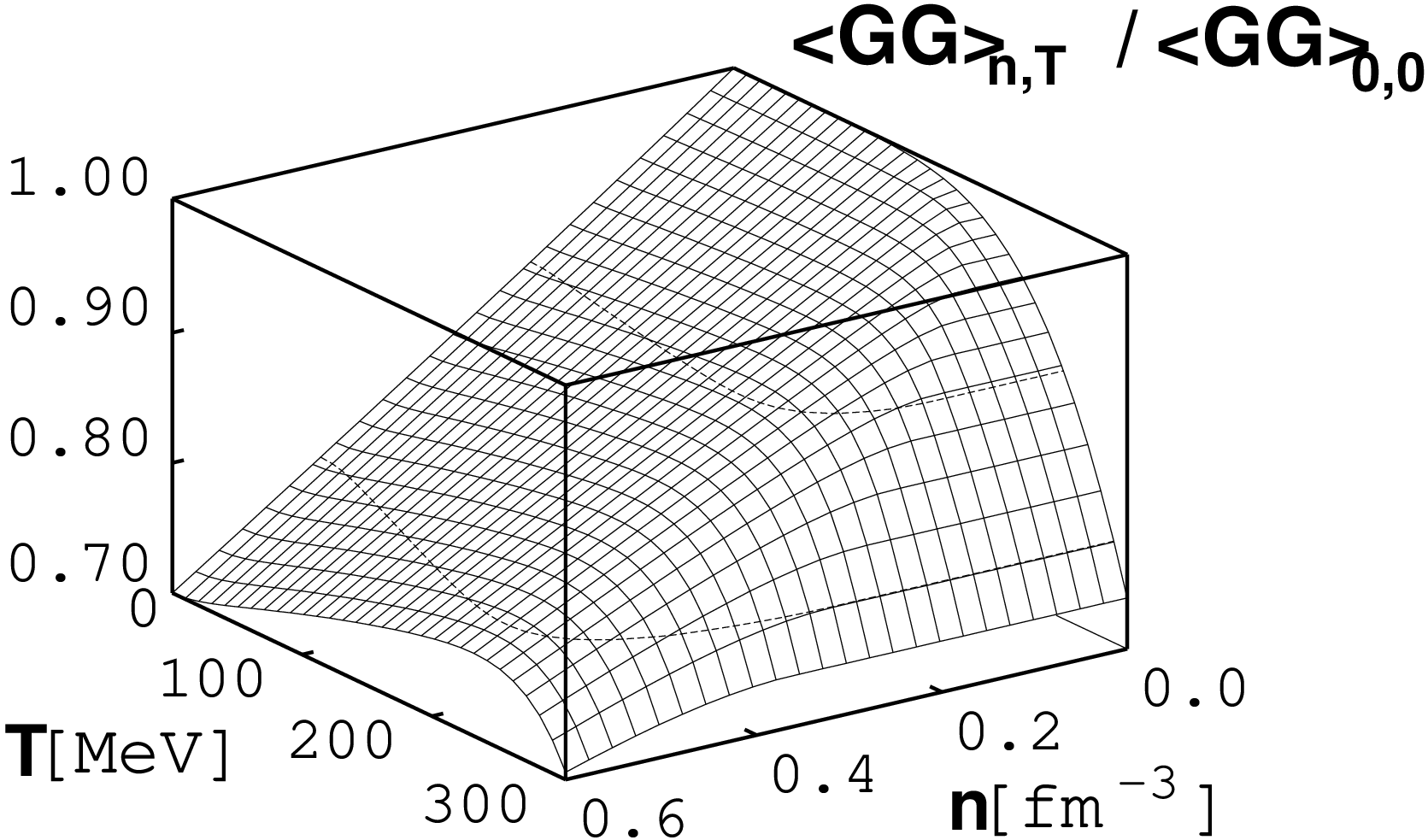,width=4cm,angle=-90}
\end{minipage}
~\vskip -3mm
\caption{{\sl The in-medium changes of condensates as a 
function of temperature and density
(left: chiral condensate, top-right: strange chiral condensate,
bottom-right: gluon condensate).
Note that the validity range of the present calculations is constrained to small
densities and temperatures. $n_N \le 3 n_0$ and $T \le 100$ MeV are
relevant for HADES.}
\label{fig_1}}
\end{figure} 

The density and temperature dependence of the
chiral condensates $\langle \bar u u \rangle$ and 
$\langle \bar s s \rangle$ and of the gluon condensate $\langle G^2 \rangle$
are exhibited in fig.~1. Note the linear density
dependence and the nearly independence of the temperature. 

The parameter $s_0^V$ in eq.~(\ref{m_V}) is fixed by requiring
maximum flatness of $m_V(M)$ within a suitably chosen Borel
window $M_{\rm min} \cdots M_{\rm max}$ determined by the 
''10\% + 50\%'' rule \cite{Leinweber}. 
If we follow \cite{Hatsuda} and choose $\kappa_N / \kappa_0 = 0.42$
and use $\kappa_0$ to adjust the vacuum values of $m_V$ to
the physical values, we get the results displayed in fig.~2.
Due to some cancellations of the coefficients entering the $\rho$ meson
sum rule, the dropping mass parameter is determined by the
four-quark condensate. For the $\omega$ meson the Landau damping
is dominating \cite{Abee}. The dropping $\phi$ meson mass
parameter reflects directly the strange chiral condensate;
a smaller value of $y$ diminishes strongly the mass shift.

\begin{figure}
\epsfig{file=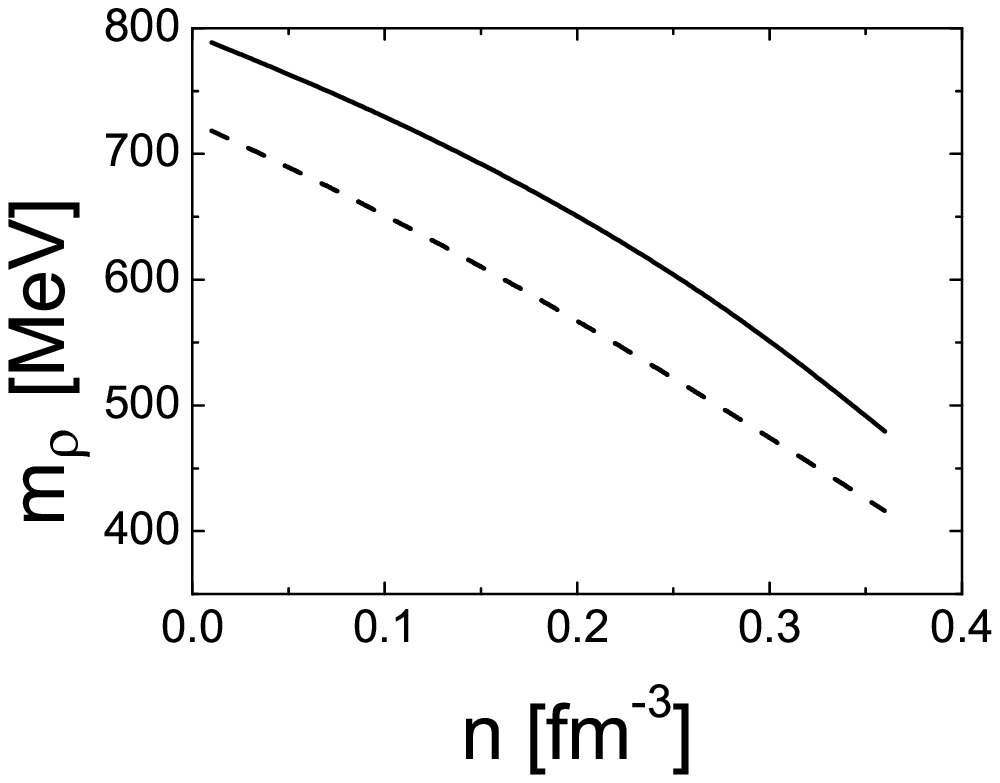,width=5cm,angle=-0}
\epsfig{file=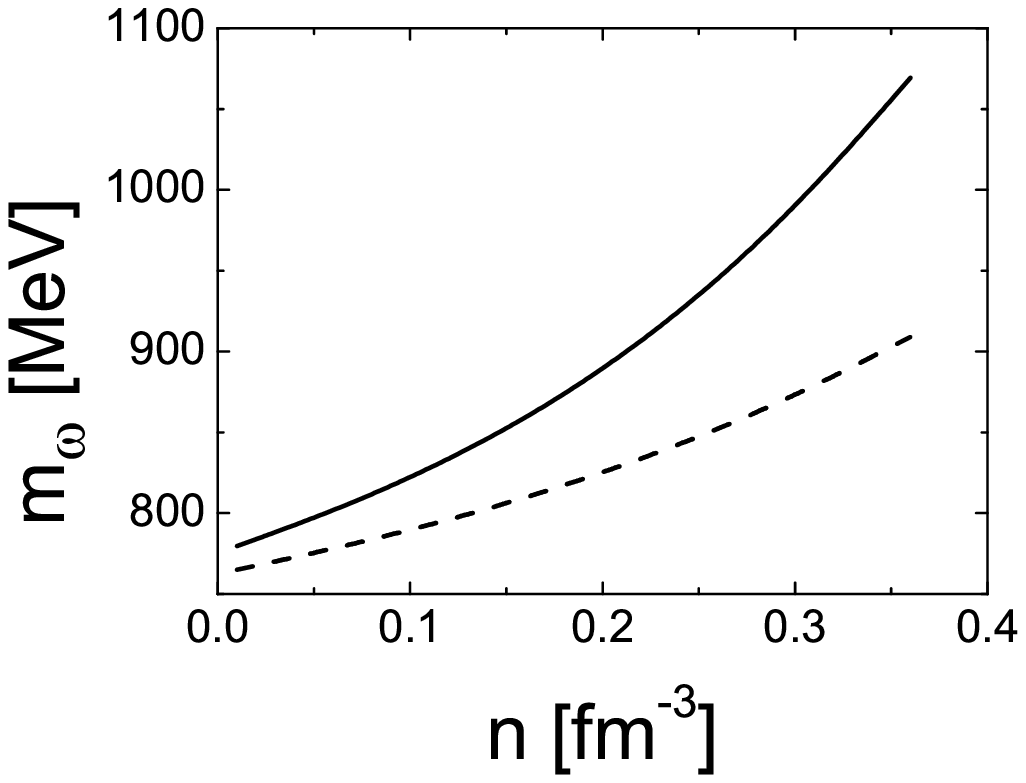,width=5cm,angle=-0}
\epsfig{file=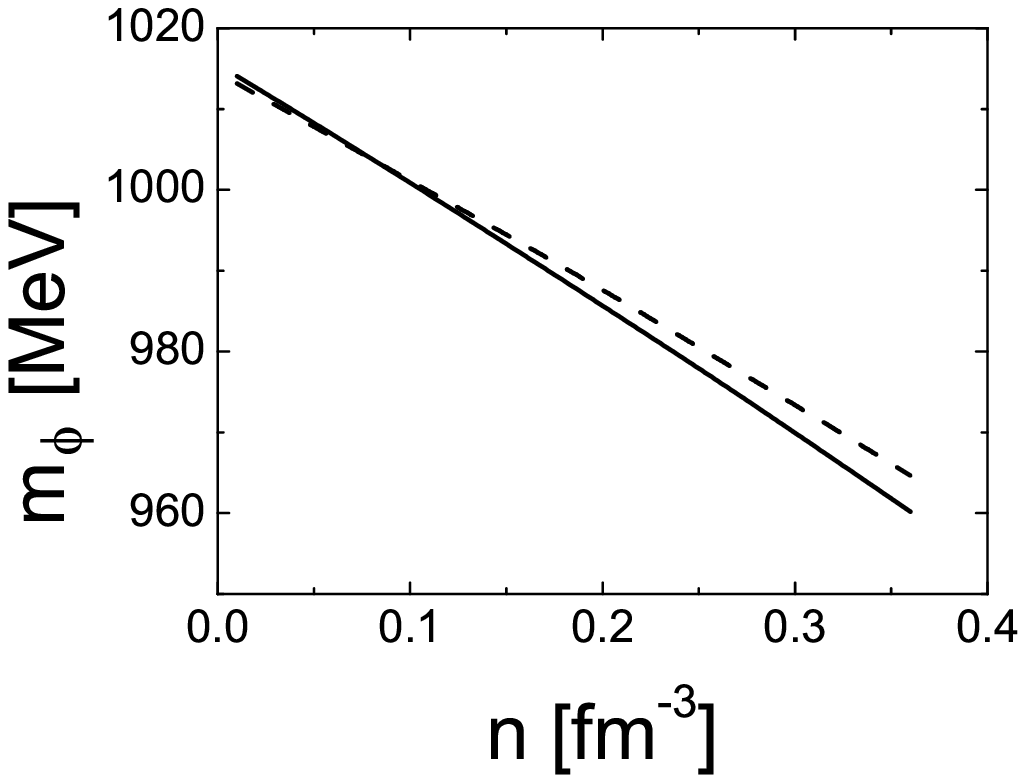,width=5cm,angle=-0}
~\vskip -5mm 
\caption{{\sl The pole mass parameters $m_{\rho, \omega, \phi}$
as a function of the density for $T = 20$ MeV (solid curves)
and 140 MeV (dashed curves) for $\kappa_N / \kappa_0 = 0.42$.}
\label{fig_2}}
~\vskip -1mm
\begin{minipage}[b]{6cm}
\epsfig{file=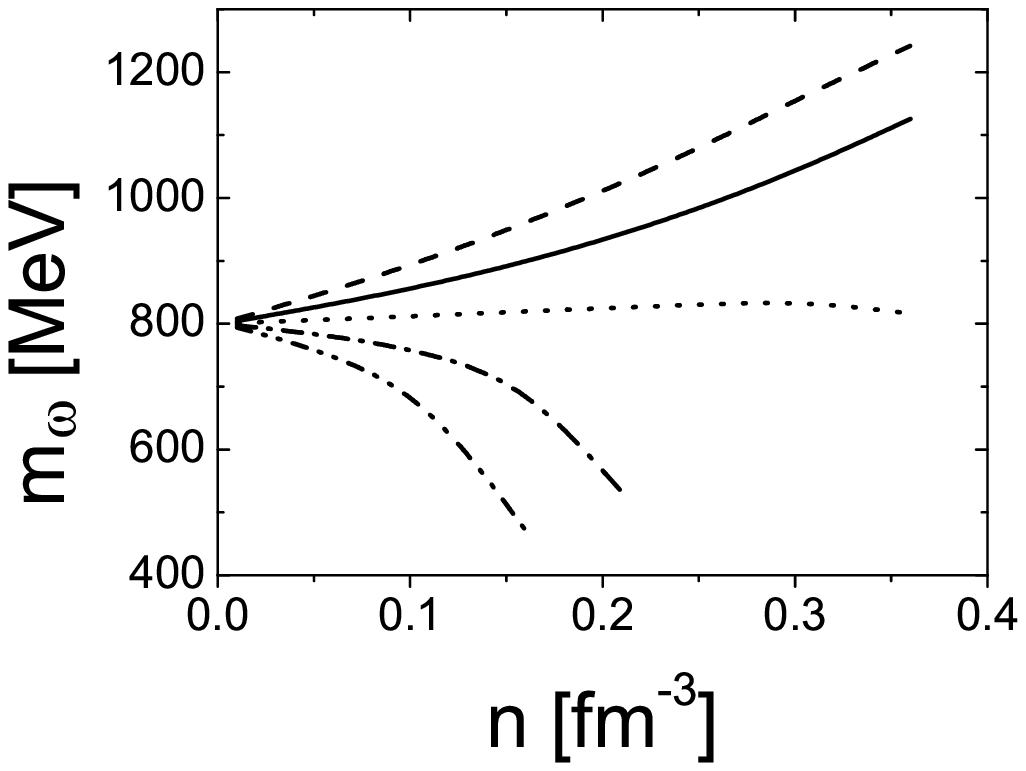,width=5.3cm,angle=-0}
\end{minipage}
\begin{minipage}[b]{8cm}
\caption{{\sl The pole mass parameter $m_\omega$
as a function of the density for $\kappa_N = 0 \cdots 4$
from top to bottom 
($T = 20$ MeV, $\kappa_0 = 3$).}
\label{fig_3}}

\vspace*{12mm}

\end{minipage}
~\vskip -3mm
\end{figure}

It turns out that the actual $\rho$ and $\omega$ mass shifts depend
on the poorly known parameter $\kappa_N$ which parameterizes
the density dependence of the four-quark condensate.
In contrast to the above ratio $\kappa_N / \kappa_0 = 0.42$
claimed in \cite{Hatsuda}, the groundstate saturation hypothesis
requires $\kappa_N / \kappa_0 = 1$.
This gives a stronger dropping $m_\rho$. For the $\omega$ meson
the rising $m_\omega$, advocated in \cite{Abee}, is changed
into a drop, as seen in fig.~3. Thus, even within the present
simple pole approximation, the qualitative behavior of the 
$\omega$ meson remains unsettled.

\section{Estimates of dilepton spectra in heavy-ion collisions}

Up to now we focused on the behavior of $m_V$ within the QSR approach
in zero-width approximation. The width-mass relation derived in \cite{Leupolt_Mosel}
is consistent with dropping $\rho$ mass as long as the width is constrained
to $\Gamma < 600$ MeV. However, a unique prediction of the in-medium modification
of the width and the mass separately seems not be possible within the QSR approach.
Rather, one has to resort to hadronic models which
predict a considerable in-medium broadening.
To elucidate the consequences of these in-medium changes together we use
the following parameterization of the emission rate of hadron matter \cite{Oleg}
\begin{equation}
\frac{dN}{d^4x \, d^4Q} = \sum_V
\frac{2 d_V}{(2 \pi)^3} \exp \left\{- \frac{u \cdot Q}{T} \right\}
\frac 1 \pi
\frac{(m_V \Gamma_V^{\rm tot})(M \Gamma_{V \to e^+ e^-})}
{(M^2 - m_V^2)^2 +(m_V \Gamma_V^{\rm tot})^2},   
\end{equation}
where $Q$ is the four-momentum vector of the lepton pair with
invariant mass $M$, $u$ denotes the four-velocity
of matter, and $d_V$ stands for the degeneracy factor of the vector meson $V$.
The corresponding invariant mass spectrum is displayed in fig.~4 for $T = 70$ MeV, 
where we assume that all vector mesons experience the overall broadening
$\Gamma_V^{\rm tot} = b \Gamma_V^{\rm tot, 0}$ with $b = 3$.
As seen in fig.~4, the $\rho$ meson is smeared out to such a degree that no
distinct peak is visible. The $\omega$ and $\phi$ can be identified
as they are still sufficiently narrow.
It is the double differential spectrum $dN / d^4 x \, dM \, dM_\perp$ which
allows access to the $\rho$ peak. As seen in the right panel of fig.~4, 
for fixed $M_\perp = 1.3$ GeV there is a pronounced $\rho$ peak with the
$\omega$ sitting on top. As known for a long time, such double differential
spectra are sensitive also to the flow of matter. This is highlighted
by a comparison of two flow velocities, 
$v = 0$ (lower curve in left panel of fig.~4) and 0.3 (upper curve).
The full invariant mass spectra are insensitive to flow effects. 

\begin{figure}
~\center
\epsfig{file=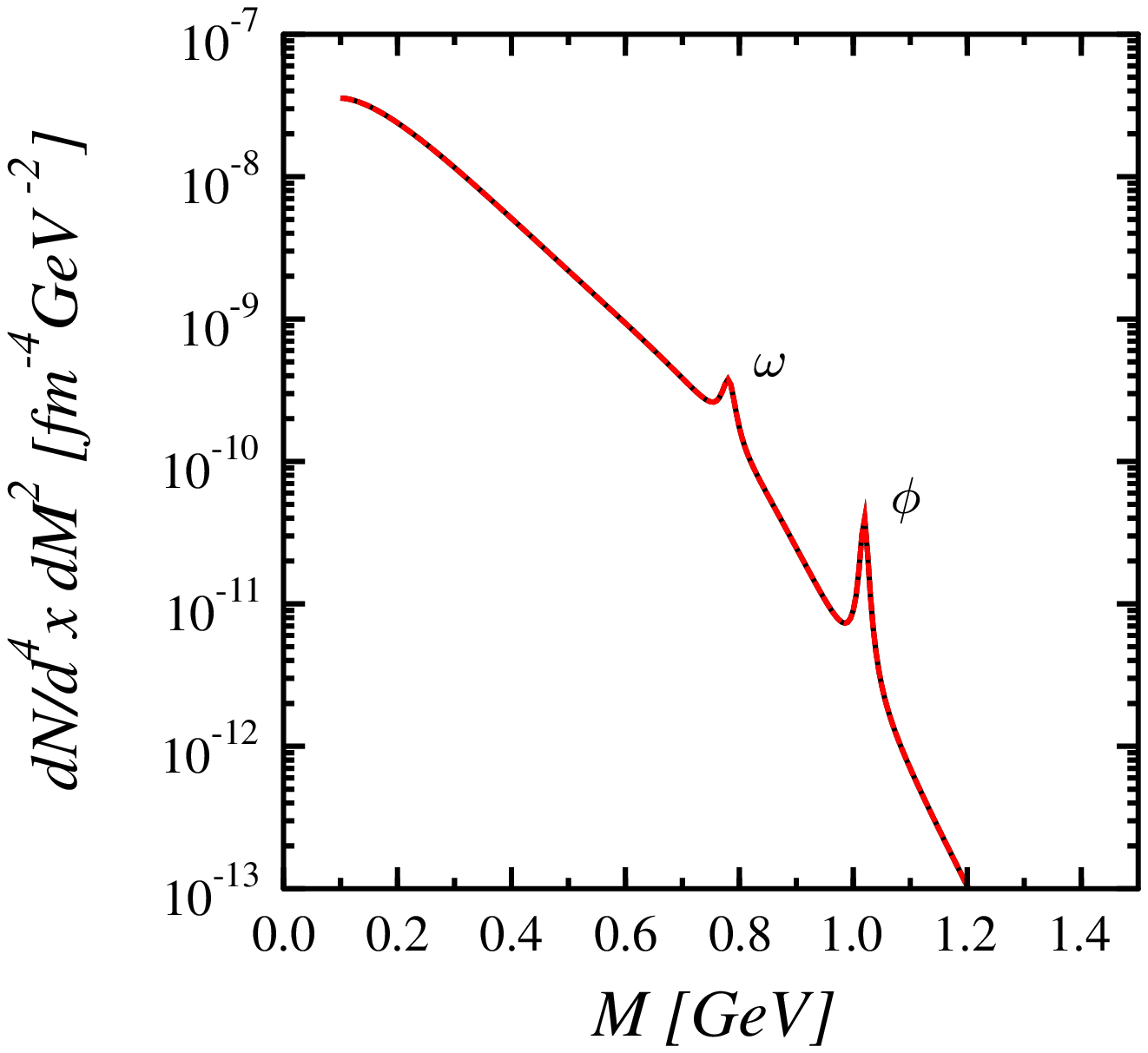,width=7cm,angle=-0}
\hspace*{6mm}
\epsfig{file=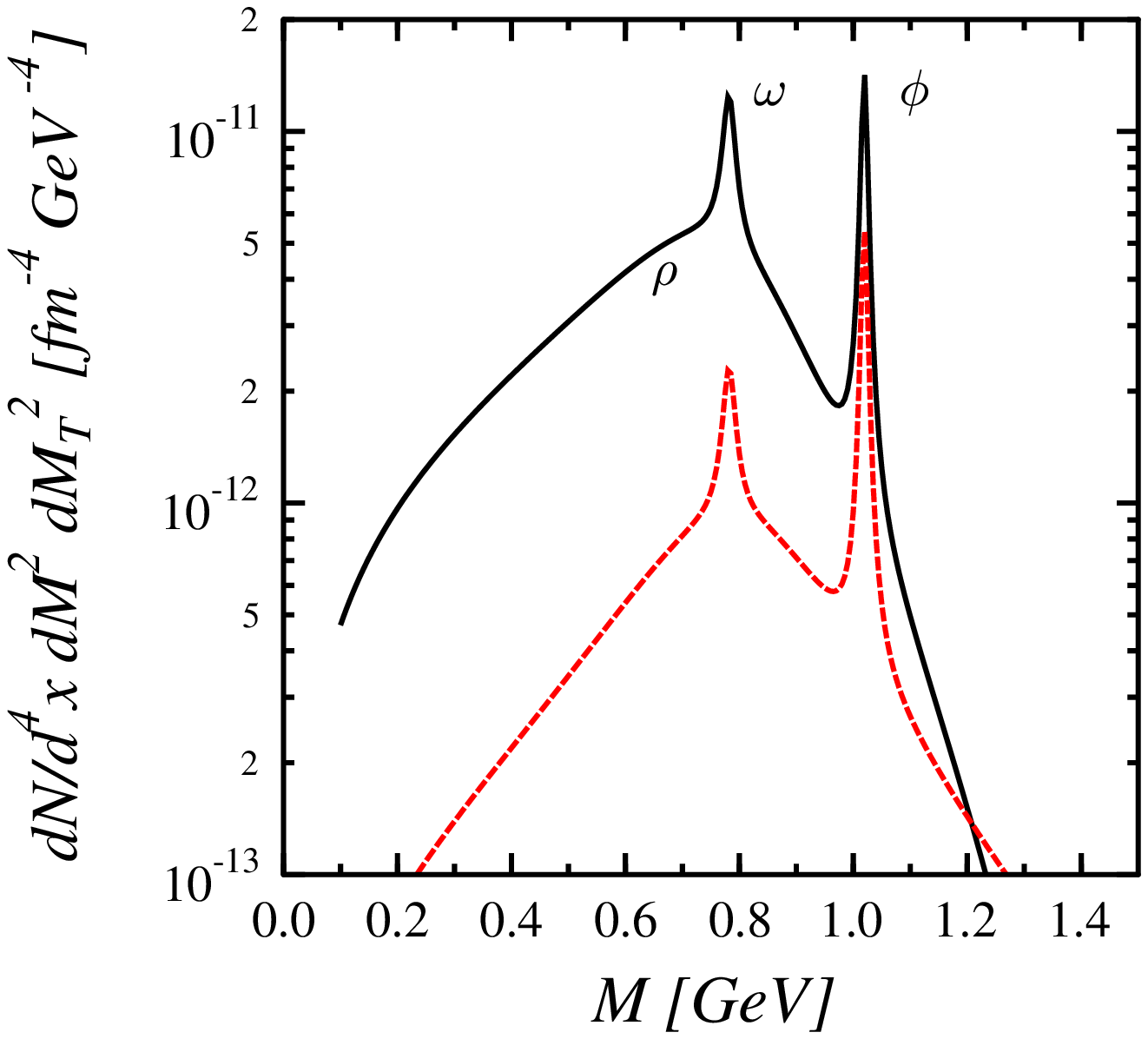,width=7cm,angle=-0}
~\vskip -4mm
\caption{{\sl Dilepton emission rate at $T = 70$ MeV 
as a function of invariant mass.
Left panel: integrated over all $M_\perp$, right panel: for $M_\perp =
1.3$ GeV
(dashed curve: $v_r = 0$, solid curve: $v_r = 0.3$).
A vector meson broadening factor of 3 is assumed.}  
\label{fig_4}}
\end{figure}

\subsection{Hydrodynamics}

In \cite{Oleg} we have studied, within a scenario of an expanding fire ball,
the resulting dilepton spectra. Instead of solving the hydrodynamical
flow equations in full detail we have used a schematic expansion dynamics
governed by  $R(t) = R_i + v_r t$ with $v_r = 0.3$, entropy conservation,
$s/n = const$, and baryon conservation ($N = 330$).
The final (freeze-out) parameters are from an analysis of hadron multiplicities
at SIS energies \cite{Redlich} $n_f = 0.3 n_0$, $T_f = 50$ MeV.
Assuming an initial density $n_i = 3 n_0$, as suggested by BUU calculations,
one gets $T_i = 90$ MeV for the initial temperature on the isentropic line.
Keeping the above assumed in-medium broadening by a factor $b = 3$ and
using the $\rho$ meson mass shift of $\Delta m_\rho = 300$ MeV
(a maximum value with respect to the above QSR studies) 
one gets the dilepton spectra
exhibited in fig.~5. According to the above QSR results for $\kappa_N \approx 2$
we employ here no mass shift of the $\omega$ meson.
While in the invariant mass spectrum again no $\rho$ peak is visible,
the double-differential spectrum with $M_\perp = 0.9$ GeV shows clearly
the shifted peak.

In \cite{Oleg} also such scenarios are studied in detail, where the density
depend broadening and mass shifts are incorporated. These studies support the need
to measure double-differential cross sections 
to have access to the $\rho$ peak.

\begin{figure}
~\center
\epsfig{file=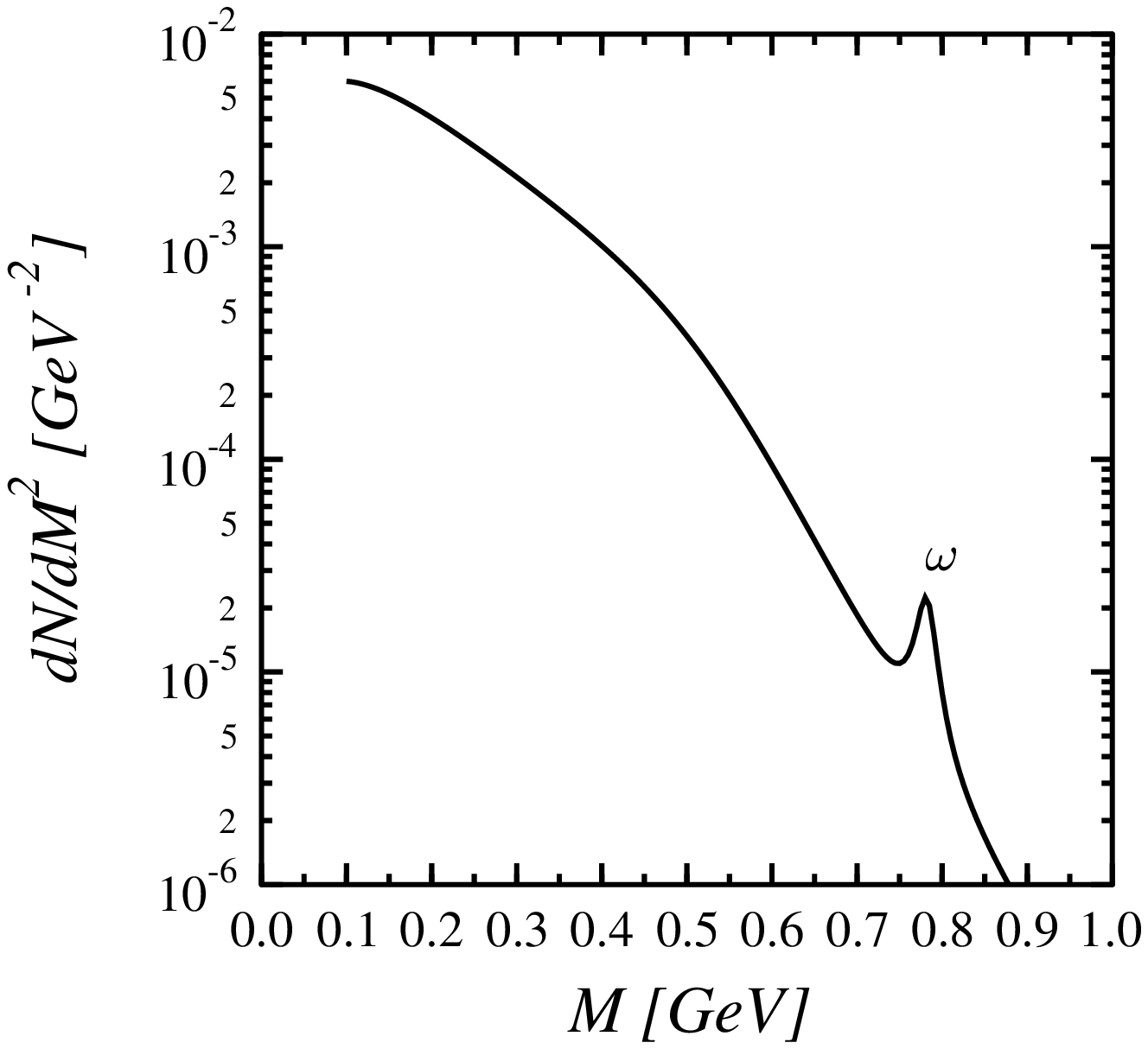,width=7cm,angle=-0}
\hspace*{6mm}
\epsfig{file=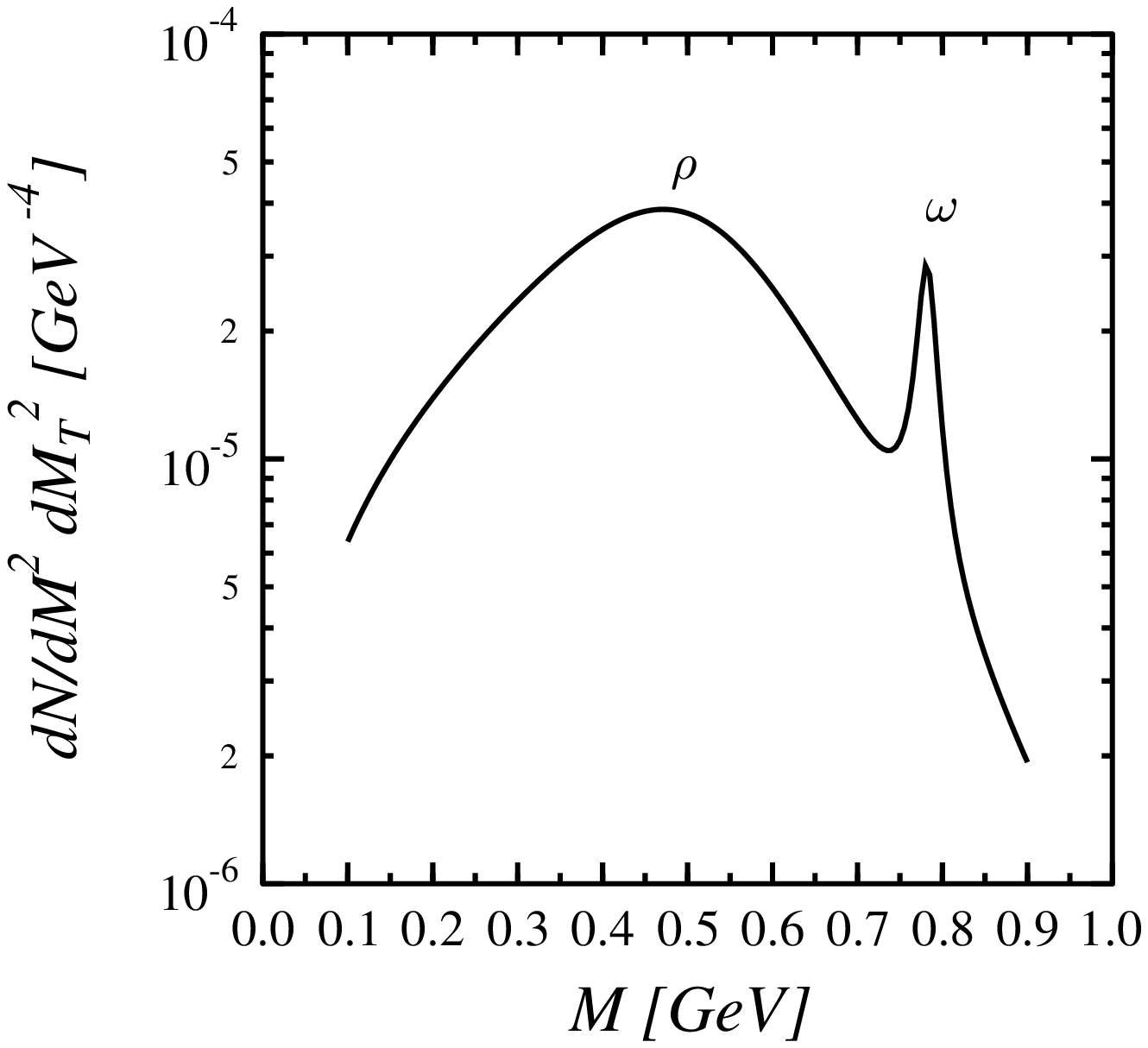,width=7cm,angle=-0}
~\vskip -4mm
\caption{{\sl Space-time integrated dilepton spectra from the expanding fire
ball described in text. Left panel: invariant mass spectrum
integrated over all $M_\perp$, right panel: $M_\perp = 0.9$ GeV. 
A vector meson broadening factor of 3 is assumed
and a $\rho$ meson mass shift of $\Delta m_\rho = 300$ MeV.}   
\label{fig_5}}
\end{figure}

\subsection{BUU calculations}

The above hydrodynamical calculations are based on local thermal and chemical
equilibrium. These stringent assumptions are not longer needed in transport
model calculations. We employ here a BUU code to calculate the dilepton
spectra for central collisions Au(1 AGeV) + Au. The results are displayed
in fig.~6 for $b = 3$ and either no mass shift (left column) or a $\rho$ mass
shift of 150 MeV (right column). 
We have included the dileptons from $\pi^+ \pi^-$ annihilations
via the intermediate $\rho$ meson, decays of $\rho$ mesons from baryon resonance
decays, $pn$ bremsstrahlung, $\Delta$ and $\eta$ Dalitz decays
and $\omega$ decays. In contrast to the above hydrodynamical calculations
the $\rho$ and $\omega$ mesons are not allowed to undergo
equilibration in the BUU code, but rather
are propagated in mean field approximation along their original trajectories.
The present BUU calculations confirm the results of the hydrodynamical model:
Only when selecting a suitable $M_\perp$ bin one gets access to the
$\rho$ peak. However, even in case of no mass shift, due to
phase space weighting the $\rho$ peak seems shifted (see left column in fig.~6). 
The role of the phase space weight is highlighted in the right column of
fig.~6, where the $\rho$ peak appears down-shifted by 
somewhat more than 150 MeV.

\begin{figure}
\epsfig{file=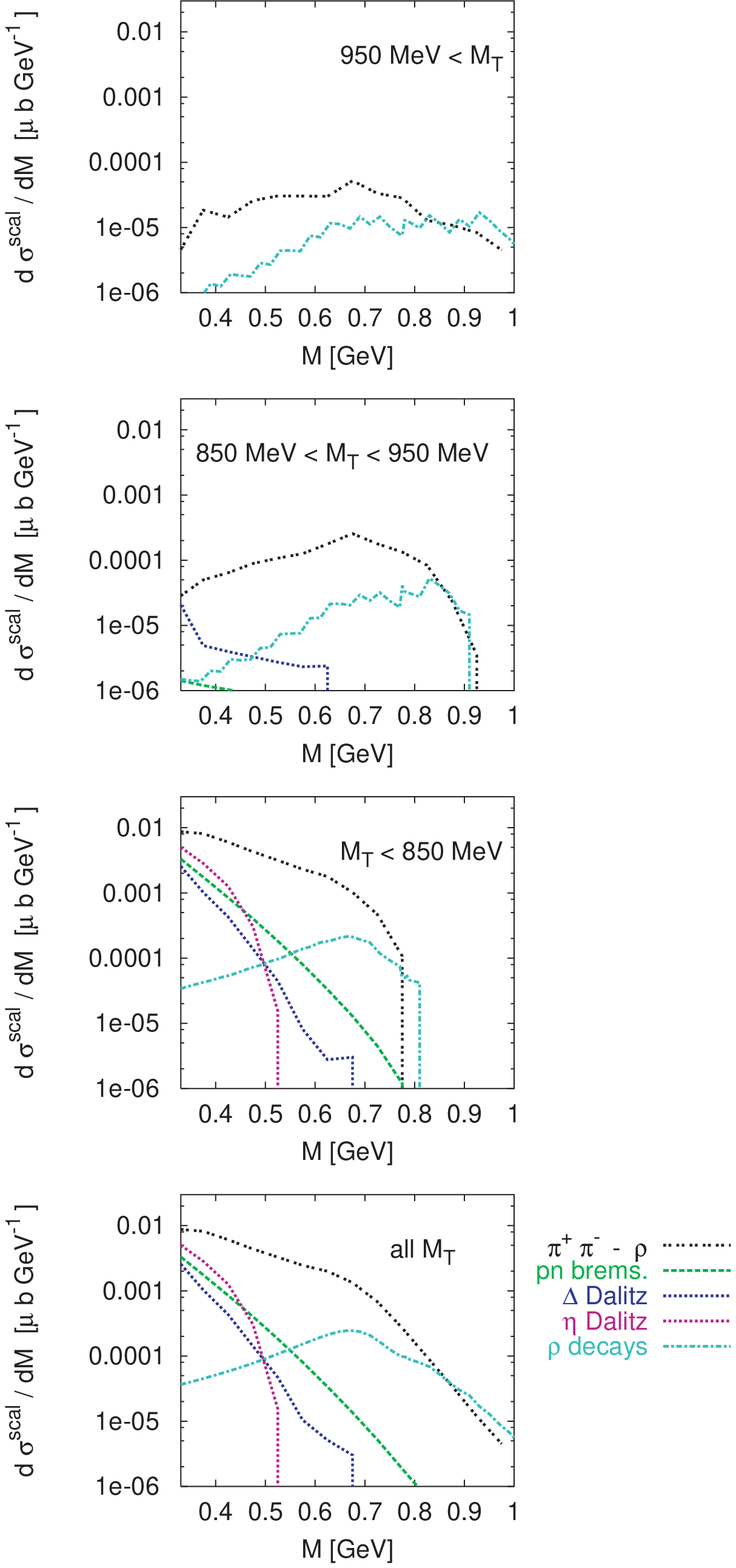,width=8.6cm,angle=-0}
\hspace*{3mm}
\epsfig{file=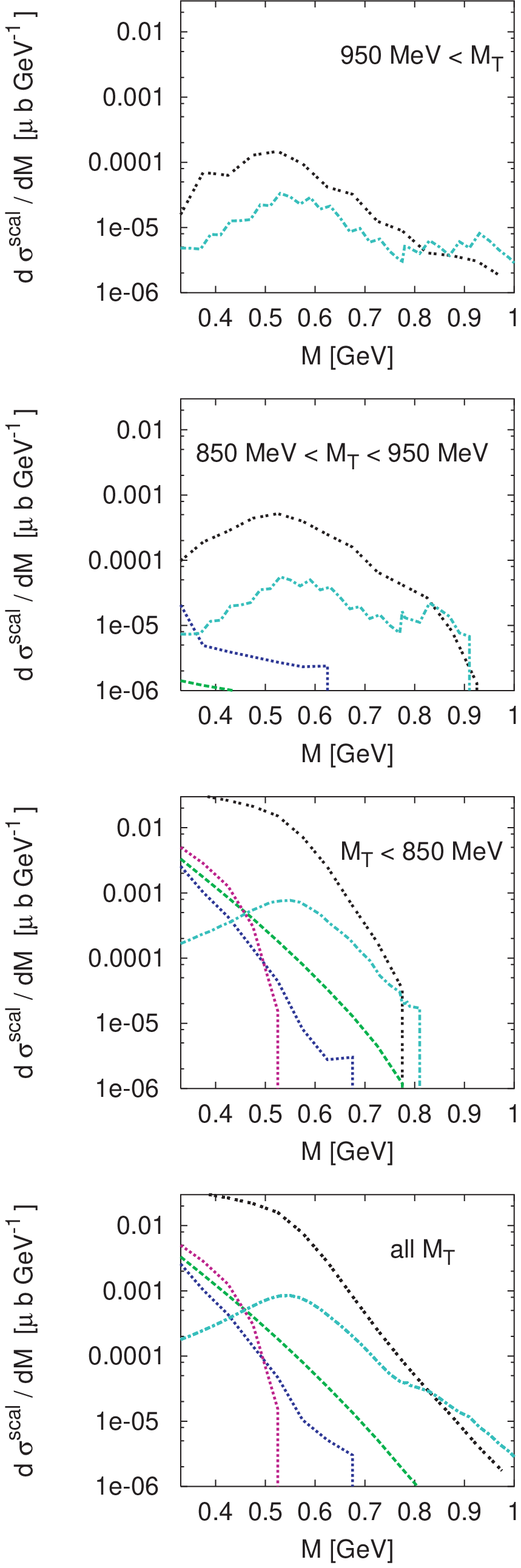,width=6.1cm,angle=-0}
\caption{{\sl Dilepton spectra as a function of invariant mass for several
$M_\perp$ bins. Left column: no mass shift,
right column: density independent mass shift $\Delta m_\rho = 150$ MeV.
An overall vector meson broadening by a factor 3 is assumed.
$\omega$ and $\phi$ contributions are not shown.
$d \sigma^{\rm scal} / dM  = \pi R^2 
(A_{\rm projetile} * A_{\rm target})^{-1} 
\int_{\Delta M_\perp} dM_\perp \, dN / (dM \, dM_\perp)$.}
\label{fig_6}}
\end{figure}

In addition, the BUU calculations show that the Dalitz and bremsstrahlung
yields do not disturb the wanted $\rho$ signal.

Various estimates demonstrate that the $\omega$ meson, even for some
broadening, remains visible as distinctive sharp peak. 
Therefore, also small shifts
should be visible. Here, however, a subtle interplay of $\omega$ decays in
medium and after freeze-out occur. To elucidate the chances to verify
possible in-medium $\omega$ modifications we have assumed
$m_\omega = m_\omega^0 - \delta m_\omega (n /n_0)$
and
$\Gamma_\omega = \Gamma_\omega^0 - \delta \Gamma_\omega (n /n_0)$ with
$\delta m_\omega = 70$ MeV and
$\delta \Gamma_\omega = 50$ MeV
and obtain the results exhibited in fig.~7.
One can see clearly a double-peak structure from vacuum decays and in-medium
decays. As expected, selecting low-$Q_\perp$ dileptons the vacuum contribution
is strongly reduced since the parent $\omega$ is nearly at rest.
The smearing of the $\omega$ due to the density dependence can mimic a
peak structure at the vacuum $\rho$ position.

\begin{figure}
\begin{minipage}[b]{8cm}
\epsfig{file=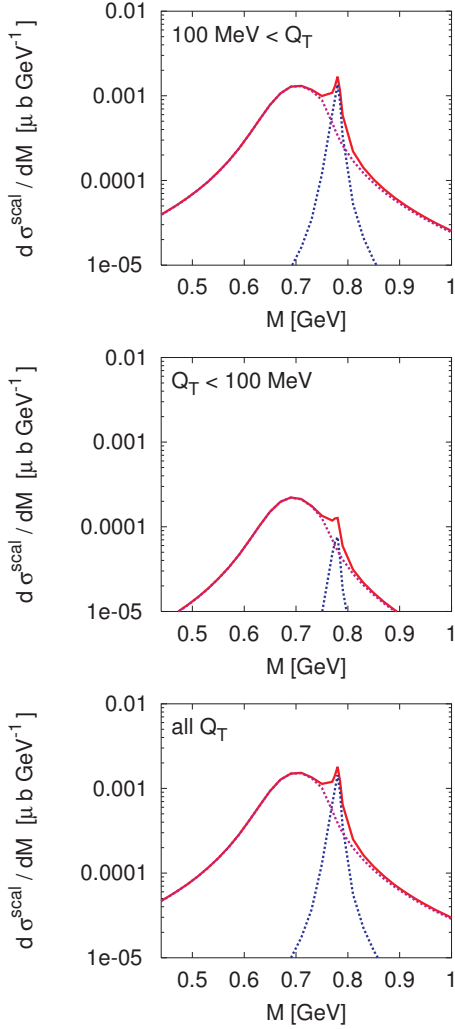,width=6cm,angle=-0}
\end{minipage}
\begin{minipage}[b]{6.5cm}
\caption{{\sl Dilepton spectra from $\omega$ decays 
as a function of invariant mass
for several $Q_\perp$ bins. The density dependent mass and width
are parameterized as 
$m_\omega = m_\omega^0 - \delta m_\omega (n /n_0)$
and
$\Gamma_\omega = \Gamma_\omega^0 - \delta \Gamma_\omega (n /n_0)$ 
with
$\delta m_\omega = 70$ MeV and
$\delta \Gamma_\omega = 50$ MeV. The sharp peak structure depicts the
vacuum decay contribution.}
\label{fig_7}}
\end{minipage}


\end{figure}


\section{Summary}

In summary we have revisited the QCD sum rule approach 
to the in-medium behavior of the light vector mesons 
in zero-width approximation.
Due to cancellations the $\rho$ meson mass shift depends 
on the four-quark condensate,
but is always decreasing with increasing density. The large Landau damping and
the uncertain four-quark condensate do not allow a definite statement on the
$\omega$ meson mass; for the ground state saturation the mass drops.
The actual mass shift of the $\phi$ meson depends sensitively on the uncertain
strangeness fraction in the nucleon; the $\phi$ mass is directly related to the
strange chiral condensate.

Our hydrodynamical fire ball calculations show that a double-differential
cross section needs to be measured to get access to the $\rho$ peak in case
of strong in-medium broadening. This is supported by transport calculations
relying on a BUU code. To amplify a possible in-medium $\omega$ shift
the selection of low-momentum pairs is very helpful.   
  

\end{document}